\documentclass[11pt,twoside]{article}
\usepackage{./asp2014}

\newcommand\fauxsc[1]{\fauxschelper#1 \relax\relax}
\def\fauxschelper#1 #2\relax{%
  \fauxschelphelp#1\relax\relax%
  \if\relax#2\relax\else\ \fauxschelper#2\relax\fi%
}
\def\Hscale{.85}\def\Vscale{.74}\def\Cscale{1.12}
\def\fauxschelphelp#1#2\relax{%
  \ifnum`#1>``\ifnum`#1<`\{\scalebox{\Hscale}[\Vscale]{\uppercase{#1}}\else%
    \scalebox{\Cscale}[1]{#1}\fi\else\scalebox{\Cscale}[1]{#1}\fi%
  \ifx\relax#2\relax\else\fauxschelphelp#2\relax\fi}

\aspSuppressVolSlug
\resetcounters

\begin{document}

\title{Radio Recombination Lines from H\,{\sc ii} Regions}
\author{Dana S. Balser,$^1$ L. D. Anderson,$^2$, T. M. Bania,$^3$
John M. Dickey,$^4$ D. Anish Roshi,$^1$ Trey V. Wenger,$^5$ and T. L. Wilson$^6$ 
\affil{$^1$National Radio Astronomy Observatory, Charlottesville, VA, USA; \email{dbalser@nrao.edu, aroshi@nrao.edu}}
\affil{$^2$West Virginia University, Morgantown, WV, USA; \email{loren.anderson@mail.wvu.edu}}
\affil{$^3$Boston University, Boston, MA, USA  ; \email{bania@bu.edu}}
\affil{$^4$University of Tasmania, Hobart, TAS, Australia; \email{John.Dickey@utas.edu.au}}
\affil{$^5$University of Virginia, Charlottesville, VA, USA  ; \email{tvw2pu@virginia.edu}}
\affil{$^6$Max-Planck-Institut f\"{u}r Radioastronomie, Bonn, Germany; \email{twilson@mpifr-bonn.mpg.de}}
}

\paperauthor{Dana S. Balser}{dbalser@nrao.edu}{0000-0002-2465-7803}{National Radio Astronomy Observatory}{}{Charlottesville}{VA}{22903}{USA}
\paperauthor{L. D. Anderson}{loren.anderson@mail.wvu.edu}{}{West Virginia University}{Department of Physics and Astronomy}{Morgantown}{WV}{26506}{USA}
\paperauthor{T. M. Bania}{bania@bu.edu}{}{Boston University}{Department of Astronomy}{Boston}{MA}{02215}{USA}
\paperauthor{John M. Dickey}{John.Dickey@utas.edu.au}{}{University of Tasmania}{School of Physical Sciences}{Hobart}{TAS}{7001}{Australia}
\paperauthor{D. Anish Roshi}{aroshi@nrao.edu}{}{National Radio Astronomy Observatory}{}{Charlottesville}{VA}{22903}{USA}
\paperauthor{Trey V. Wenger}{tvw2pu@virginia.edu}{}{University of Virginia}{Department of Astronomy}{Charlottesville}{VA}{22904}{USA}
\paperauthor{T. L. Wilson}{twilson@mpifr-bonn.mpg.de}{}{Max-Planck-Institut f\"{u}r Radioastronomie}{}{Bonn}{}{D-53121}{Germany}


\begin{abstract}

  The ngVLA will create a Galaxy-wide, volume-limited sample of
  H\,{\sc ii} regions; solve some long standing problems in the
  physics of H\,{\sc ii} regions; and provide an extinction-free star
  formation tracer in nearby galaxies.

\end{abstract}

\section{Introduction}

Here we discuss three scientific areas that could be revolutionized by
ngVLA observations of radio recombination line (RRL) and continuum
emission from H\,{\sc ii} regions: Galactic structure, H\,{\sc ii}
region physics, and star formation in nearby galaxies.  

H\,{\sc ii} regions are zones of ionized gas surrounding
recently-formed high-mass OB-type stars.  They are the archtypical
tracer of spiral structure.  Since they are young ($< 10$\,Myr), their
chemical abundances represent Galactic abundances today, and reveal
the effects of billions of years of Galactic chemical evolution.
H\,{\sc ii} regions are often part of a high-mass star formation
complex that consists of a thin neutral photodissociation region (PDR)
between a hot ($\sim 10^{4}$\,K) H\,{\sc ii} region and dense, cool
($\sim 10-100$\,K) molecular material from which the stars were
formed.

H\,{\sc ii} regions are the brightest objects in the Milky Way at
infrared and mm wavelengths and can be detected across the Galaxy.
They are characterized in the infrared by $\sim 20\,\mu$m emission
from stochastically heated small dust grains mixed with the ionized
gas surrounded by $\sim 10\,\mu$m emission from polycyclic aromatic
hydrocarbon (PAH) molecules located in the PDR \citep{anderson14}.
The ionized gas produces bright, free-free (thermal) radio continuum
emission that is coincident with the $\sim 20\,\mu$m infrared
emission.  When free electrons recombine and cascade to the ground
state they produce recombination line emission which is found at radio
wavelengths for the higher principal quantum number levels (${\rm n} >
50$).  Currently, only emission from hydrogen and helium is detected
in H\,{\sc ii} regions at radio wavelengths, but heavier elements
(e.g., carbon) are detected in the denser, cooler PDRs
\citep{wenger13}.

RRLs are an excellent, extinction-free diagnostic of H\,{\sc ii}
regions.  Unlike radio continuum emission, they uniquely probe the
thermal emission and therefore unambiguously identify an H\,{\sc ii}
region.  The RRL parameters alone provide information about H\,{\sc
  ii} region kinematics \citep{anderson12}, turbulent motions
\citep{roshi07}, and He/H abundance ratios \citep{balser06}.
Measuring the RRL and continuum emission together allows the
calculation of many physical H\,{\sc ii} region properties: electron
(thermal) temperature, rms electron density, the H-ionizing stellar
luminosity (number of OB-type stars), etc. \citep{gordon09}.

The main limitation of using RRLs to probe Galactic structure and star
formation compared to similar tracers at optical or infrared
wavelengths is that they have much lower intensities.  This is
particularly an issue in external galaxies.  For example, RRL emission
has been primarily detected from nearby starburst galaxies powered by
massive young star clusters.  Recently, however, RRL emission was
detected in the normal galaxies M51 and NGC\,628 at centimeter
wavelengths \citep{luisi18}.  Moreover, for some radio facilities the
spatial resolution is not sufficient to resolve individual H\,{\sc ii}
regions, which limits the analysis.  The ngVLA will provide both the
sensitivity and spatial resolution to remedy these problems and place
RRLs as an H\,{\sc ii} region diagnostic on par with, for example,
H$\alpha$.

\section{Galactic Structure}

The ngVLA is necessary to create a Galaxy-wide, volume-limited sample
of H\,{\sc ii} regions.  Since radio waves penetrate the dust located
within the Galactic disk, an H\,{\sc ii} region map of the entire
Milky Way is possible at these wavelengths.  This is needed to
understand the global properties of the Milky Way, and to compare its
star formation to that of external galaxies.  There is strong evidence
that we have only discovered about 25\% of the H\,{\sc ii} regions in
the Milky Way, which impedes our understanding of Galactic structure
and Galactic chemical evolution \citep{anderson14}.

The implications for Galactic structure studies using RRLs were
recognized shortly after their discovery in 1965 and prompted
large-scale H\,{\sc ii} region surveys \citep{wilson70,
  reifenstein70}.  RRLs measure the systemic velocity of an H\,{\sc
  ii} region and therefore their motion around the center of the
Galaxy.  Using a rotation curve model together with the velocity
yields the Heliocentric distance of the H\,{\sc ii} region.  Distances
computed in this way are known as ``kinematic distances.''
Improvements in receiver performance resulted in follow-up surveys in
the 1980s that increased the number of sources \citep{wink83,
  caswell87, lockman89}.  These RRL surveys, however, were limited in
sensitivity and had inaccurate distance determinations.  A third
generation of Galactic H\,{\sc ii} region discovery surveys (HRDS) is
almost complete using primarily the Green Bank Telescope (GBT) in the
Northern sky \citep{bania10} and the Australia Telescope Compact Array
(ATCA) in the Southern sky \citep{brown17}.  The HRDS has doubled the
number of known H\,{\sc ii} regions.  It has the sensitivity to detect
all nebulae ionized by an O-type star out to a distance of 20\,kpc
from the Sun.  This has primarily been achieved by the flexibility of
spectrometers that can simultaneously observe multiple RRLs which can
be averaged together to increase the signal-to-noise ratio (SNR).
Maser parallax distances together with better models of Galactic
rotation have improved our understanding of kinematic distance
determinations \citep{reid14, wenger18}.

These new surveys have made several discoveries about the
morphological and chemical structure of the Milky Way.  {\it Cold H}
\textit{\fauxsc{i}} {\it gas and molecular clouds are often used to
  define spiral structure but such tracers are not always associated
  with high-mass star formation} \citep{anderson09}.  Therefore,
\citet{koo17} used H\,{\sc i} surveys together with the HRDS to
characterize the spiral structure in the outer Galaxy and found that a
four-arm spiral model with pitch angle of $12^\circ$ was a good fit to
the data.  Yet the number of H\,{\sc ii} regions sampling the
outer-most spiral arms is limited.  For example, deep RRL observations
with the GBT were only able to detect a handful of H\,{\sc ii} regions
in the outer Scutum-Centaurus arm \citep{armentrout17}.  (N.B., the
RRL spectral sensitivity of the Jansky Very Large Array (JVLA) is
comparable to the GBT.)

For H\,{\sc ii} regions in thermal equilibrium the nebular electron
temperature is determined by the abundance of the coolants (O, N, and
other heavy elements). Balser et al. (2015) used this to derive
electron temperatures for a sub-set of the HRDS H\,{\sc ii} regions
and used them to derive the nebular [O/H] abundances.
Figure~\ref{fig:metals} shows the resulting metallicity map and
reveals the chemical structure of the Milky Way disk.  The radial
metallicity gradient is obvious from this map and has been detected
before, but there is also azimuthal structure that was revealed for
the first time.  This may be due to radial mixing from the Galactic
Bar whose major axis is aligned toward an azimuth of about
$30^{\circ}$.  The sample is limited, however, since only the most
accurate RRL and continuum data is selected.  More sources are
required to better characterize this chemical structure and therefore
constrain the physical mechanisms that might be in play.  The ngVLA
will add thousands of sources to Figure~\ref{fig:metals} (see below)
and reveal any detailed metallicity structure across the Galactic
disk.

\articlefigure{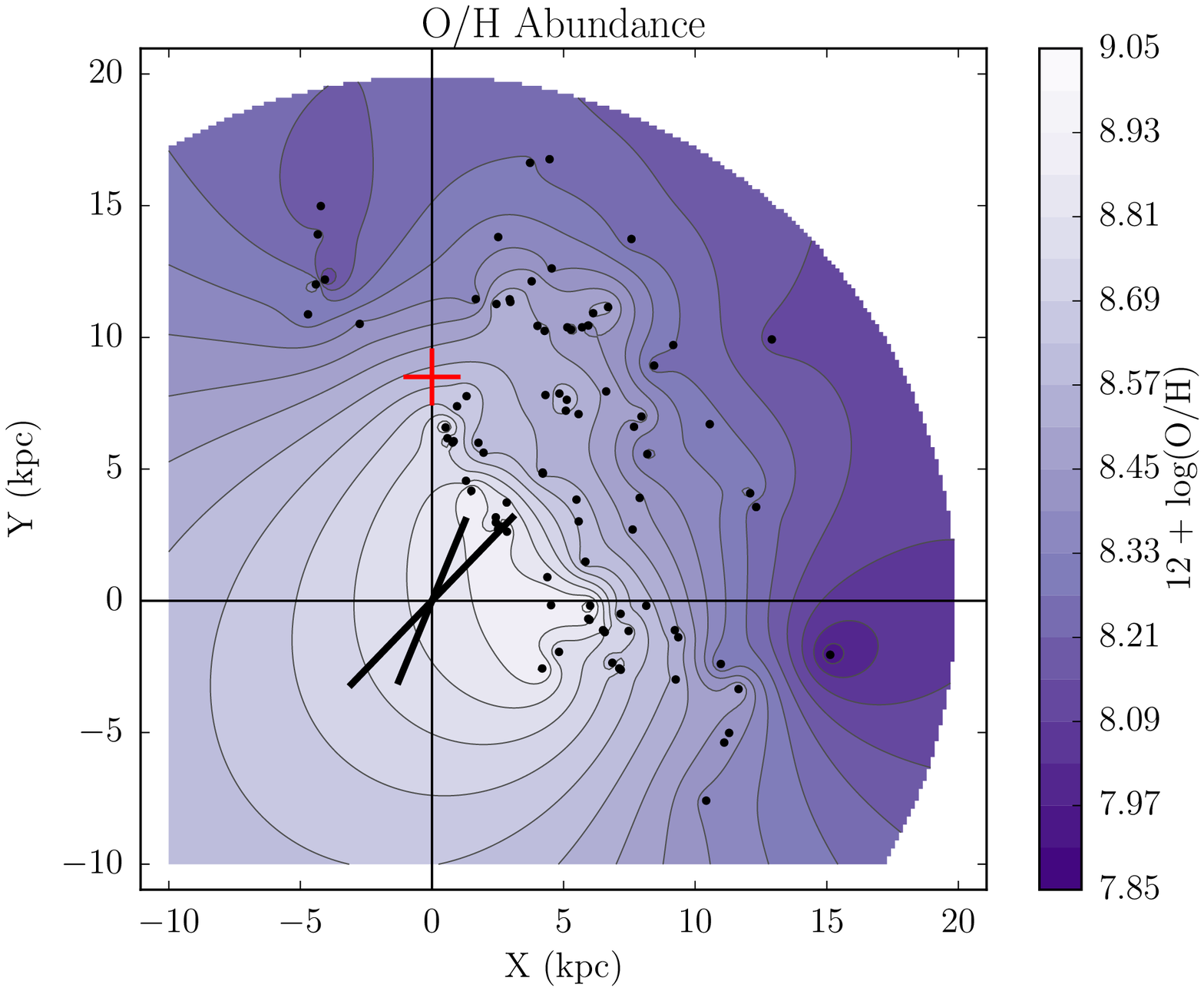}{fig:metals}{Face-on Galactic [O/H]
  abundance ratio image using Kriging to interpolate between the
  discrete H\,{\sc ii} region values taken from \citet{balser15}.  The
  points indicate the location of the discrete H\,{\sc ii}
  regions. The solid lines intersect at the Galactic Center. The red
  lines mark the location of the Sun. The thick lines correspond to
  the central locii of the putative ``short'' and ``long'' bars.}

The {\it WISE} catalog of Galactic H\,{\sc ii} regions contains about
8000 objects identified by their mid-infrared and radio properties as
H\,{\sc ii} region \citep{anderson14}.  Currently only $\sim 2000$
sources in this catalog are known H\,{\sc ii} regions with RRL or
H$\alpha$ emission.  Of the remaining sources $\sim 2000$ have
spatially coincident radio continuum emission and are very likely
H\,{\sc ii} regions, and $\sim 4000$ are radio quiet at the
sensitivity limits of existing radio continuum surveys.  There is some
evidence that the radio quiet sources in the {\it WISE} H\,{\sc ii}
region catalog are bona fide H\,{\sc ii} regions (Armentrout et al.
2018, in prep.).  Because of their extreme luminosities, H\,{\sc ii}
regions are essential for computing extragalactic star formation rates
in combination with infrared and other radio data.  The HRDS, when
complete, will be the first survey to detect all H\,{\sc ii} regions
created by at least one O-type star, but there will remain thousands
of likely H\,{\sc ii} regions whose RRL emission cannot be detected
with extant facilities.

The ngVLA will have the sensitivity to create a Galaxy-wide,
volume-limited sample of H\,{\sc ii} regions.  The optimal frequency
range is 4-12\,GHz (ngVLA band 2), where classical H\,{\sc ii} regions
are optically thin, in LTE, and many adjacent transitions can be
averaged to increase the SNR. The HRDS spectral sensitivity at these
frequencies is $\sim 1\,$mJy.  To detect all sources in the {\it WISE}
Galactic H\,{\sc ii} region catalog requires a spectral sensitivity of
$\sim 3\,\mu$Jy \citep{anderson14}.  This assumes a radio continuum
flux density of 0.2\,mJy at 21\,cm, a SNR of 5, an optically thin
nebula, and a line-to-continuum ratio of 0.1 at 8\,GHz
\citep[e.g.,][]{brown17}.  Using the ngVLA spectral sensitivity from
\citet{selina17} for a 50\,mas beam\footnote{The ngVLA spectral rms is
  $37.3\,\mu$Jy\,beam$^{-1}$ for a beam size of 50\,mas at 8\,GHz with
  an integration time of 1\,hr and a FHWM line width of
  10\,km\,s$^{-1}$.}, the weakest sources in the {\it WISE} catalog
would require 15\,hr of integration time.  The ngVLA sensitivity will
be less for sources that are resolved by the longest baselines.  At a
distance of 20\,kpc an H\,{\sc ii} region with a physical size of
1\,pc has an angular size of 10$^{\prime\prime}$.  Since the ngVLA
core of antennas will contain a large fraction of the collecting area
the sensitivity should not be significantly reduced.  In principle,
the GBT could fill in the missing flux density.  Phased array feeds in
the central ngVLA core would increase the efficiency.  Such a survey
would complement MeerGAL, a high frequency ($\sim 10\,$GHz) survey of
the Galactic plane with MeerKat in the Southern sky.

A Galactic H\,{\sc ii} region pointed survey toward {\it WISE} H\,{\sc
  ii} region candidates may be possible with the {\it James Webb Space
  Telescope} (JWST), which will be sensitive to wavelengths from
$0.6-28\micron$, using collisonally excited lines (CELs) in the
mid-infrared.  RRLs provide a more direct measure of the ionizing
flux, but CELs probe the hardness of the radiation field.  Most
optical and near-infrared studies, however, trace the kinematic and
chemical structure of stars.  Recent surveys such as APOGEE, GAIA-ESO,
LAMOS, and RAVE are sampling hundreds of thousands of stars and will
soon have accurate distances from GAIA.  But these surveys do not
probe the gas component of the Milky Way or very far into the disk due
to extinction by dust.  Understanding the morphological and chemical
structure of both the stars {\bf and} the gas is critical to constrain
Galactic formation and evolution models \citep[e.g.,][]{kubryk13}.

\section{H\,{\sc ii} Region Physics}

The ngVLA could solve some long standing problems in the physics of
H\,{\sc ii} regions including, for example, constraining the collision
rates from free electrons, characterizing temperature fluctuations,
and understanding the role of magnetic fields within H\,{\sc ii}
regions.

Recombination lines from hydrogen and helium at radio wavelengths are
strongly affected by collisions from free electrons and therefore
accurate collision rates are important for predicting the line
opacity.  Under some physical conditions the opacities can vary by as
much as a factor of five depending on the rates used
\citep[F. Guzm\'{a}n, private comm.; also see][]{guzman16}.  Accurate
measurements of the RRL intensity ratio over a frequency range wide
enough for the opacity to vary significantly would provide vital
observational constraints for the collision rates.  For example, the
flexible spectrometer on the ATCA allows RRLs to be observed
simultaneously from 6-12\,GHz so that we can begin to constrain the
atomic physics (see Figure~\ref{fig:shrds}).  We are working on a
detailed strategy to detect each RRL transition with a high SNR over a
variety of physical conditions. This will require the sensitivity of
the ngVLA.

\articlefiguretwo{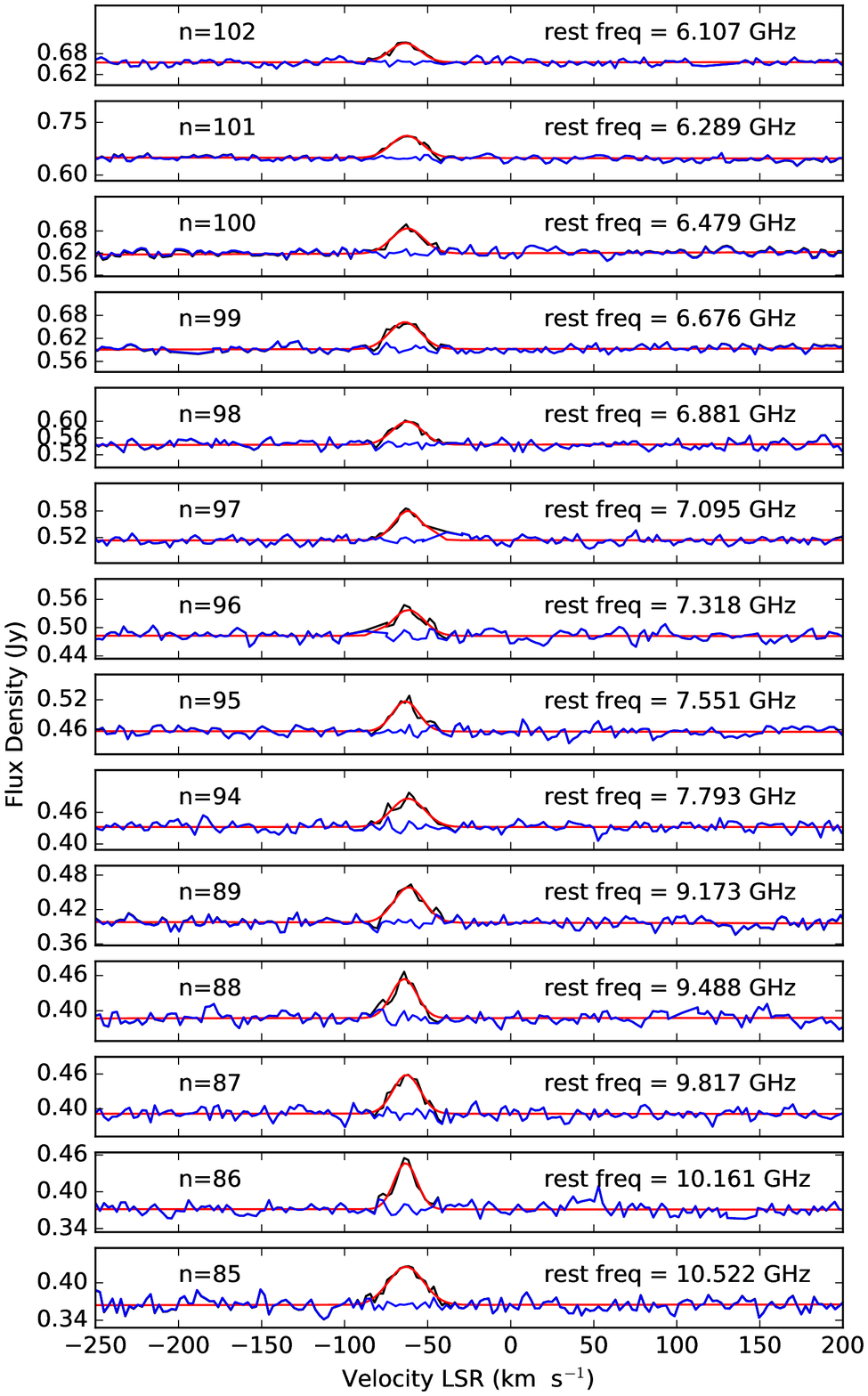}{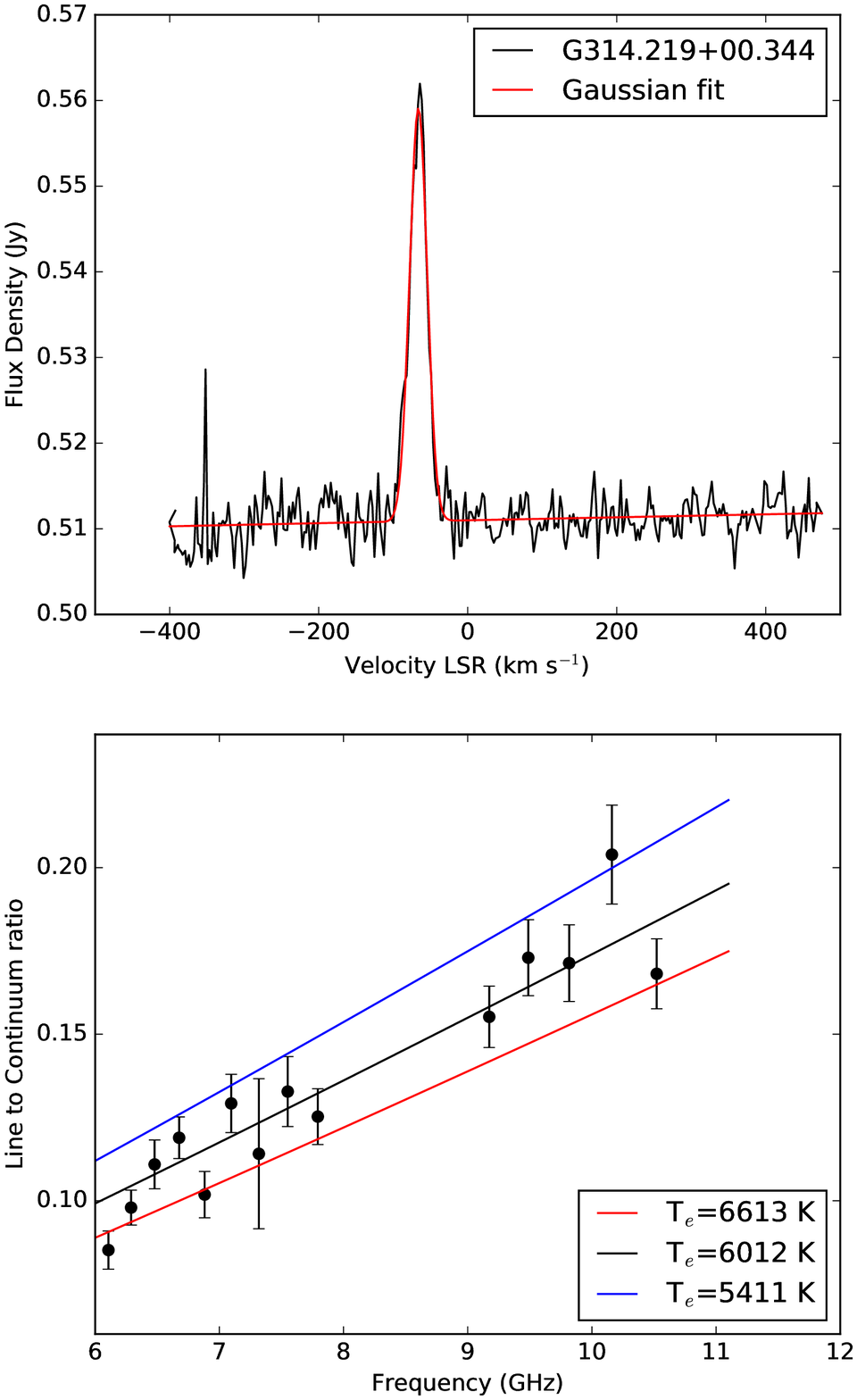}{fig:shrds}{RRL spectra for
  G314.219+00.344 using the ATCA taken from
  \citet{brown17}. Individual spectra (left) and an averaged spectrum
  (upper-right) are shown where the red curves are Gaussian fits to
  the data.  The lower-right panel plots the line-to-continuum ratio
  as a function of frequency.}

One of the key problems in the physics of ionized nebulae is that
abundances derived from recombination lines (RLs) are significantly
larger, by as much as a factor of two, than those derived from CELs
\citep{garcia-rojas07}.  These differences can be explained by
temperature fluctuations in the ionized gas.  The concept of
temperature fluctuations was developed by \citet{peimbert67} to
explain how electron temperatures derived from RLs were different than
those derived from CELs.  Temperatures are also expected to increase
at the H\,{\sc ii} region boundary because of photon hardening.  That
is, photons with energies of 20\,eV have a smaller absorption cross
section so they can reach the nebular boundary where they are absorbed
and thus raise the electron temperature.  \citet{wilson15} suggested
that optical observations produced different electron temperature
profiles in the Orion nebula compared with RRLs due to the effects of
dust.  The radio observations were performed with the GBT instead of
the JVLA because Orion is extended, but to probe small-scale structure
requires better spatial resolution.  The smallest scales over which
the temperatures vary is not known.  Therefore to measure both the
small and large-scale temperature variations in H\,{\sc ii} regions
requires calculating the radio line-to-continuum ratio over a wide
range of spatial scales with high sensitivity.  This may be possible
by combing data from the JVLA and GBT, but the systematic errors using
this method may be too large.  The internal calibration with the ngVLA
would provide a more accurate measure of the electron temperature over
a wider range of spatial scales with significantly better sensitivity.

Magnetic fields may play an important role in the formation of stars
and the Zeeman effect is the only method that directly measures the
{\it in-situ} magnetic field strength, but it is a challenging
experiment.  Most of the work thus far has probed molecular (e.g., CN)
or neutral (e.g, H\,{\sc i}) components of high-mass star formation
regions \citep{crutcher10}.  To our knowledge there has never been a
Zeeman detection from RRLs \citep[e.g.,][]{troland77}, so an ngVLA
detection would be the first.  Such a measurement is difficult since
the hydrogen RRL width ($\sim 25\,$km\,s$^{-1}$) is broader than the
Zeeman splitting and therefore a very high SNR is required.  High
spatial resolution is necessary since magnetic field variations have
been detected and will be averaged over a large beam.  Since many
bright H\,{\sc ii} regions are extended, however, the larger spatial
scales, or shorter baselines for an interferometer, must be sampled.

We estimate the sensitivity of the ngVLA for Zeeman measurements in
H\,{\sc ii} regions by assuming a hydrogen RRL intensity of
50\,mJy\,beam$^{-1}$ and a line width of 30\,km\,s$^{-1}$ observed
with a resolution of 1\,arcsec at 8\,GHz.  Using the ngVLA spectral
sensitivity from \citet{selina17} for a 1\,arcsec beam and Equation 2
from \citet{troland82}, we estimate a 3\,$\sigma$ limiting
line-of-sight magnetic field strength of 295\,$\mu$G in 1\,hr of
integration time.  This assumes we are able to average 20 RRLs with
six spectral channels across the FWHM line width.

\section{Star Formation in Nearby Galaxies}

Using RRLs the ngVLA will provide an extinction-free star formation
tracer in nearby galaxies that is comparable to H$\alpha$ for the
sensitivity and spatial resolution afforded by existing optical
instruments.  That is particularly important in extreme environments
with clusters of massive stars since star formation may proceed
differently in these objects.  These environments are, moreover, often
opaque at ultraviolet, optical, or near infrared wavelengths due to
dust.

\articlefigure{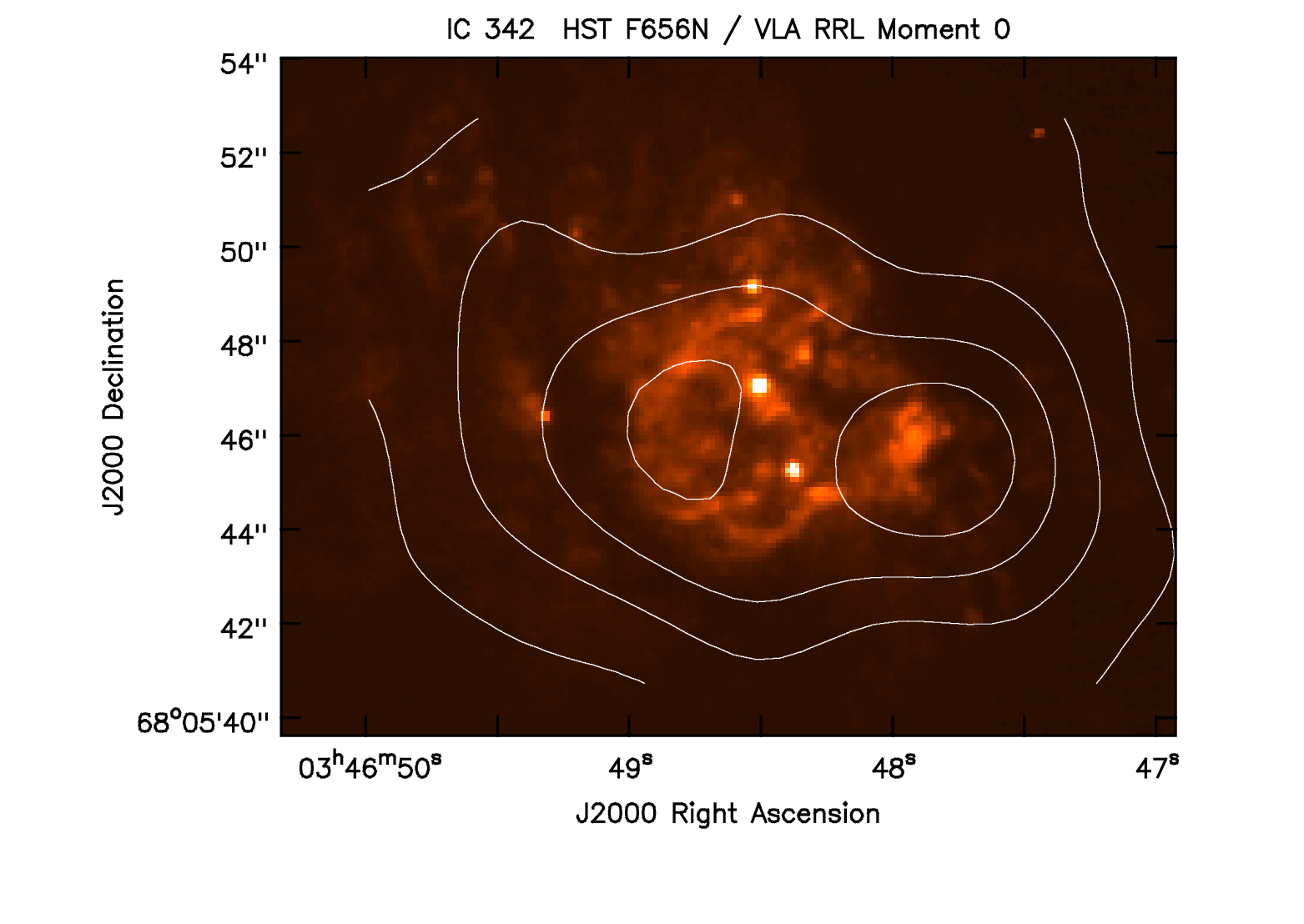}{fig:ic342}{Star formation diagnostics in the
  IC\,342 nuclear region taken from \citet{balser17}. Shown in color
  is the HST H$\alpha$ (F656N) image together with the JVLA Ka-band
  RRL-integrated intensity map in contours. The contour levels are
  0.2, 0.4, 0.6, and 0.8 times the peak intensity
  (0.1324\,Jy\,beam$^{-1}$\,km\,s$^{-1}$). The JVLA synthesized beam
  size is 4$\rlap.{^{\prime\prime}}$5 $\times$
  4$\rlap.{^{\prime\prime}}$5.}

Radio continuum measurements are excellent probes of these regions
since they can penetrate the gas and dust that surrounds the young
star clusters.  Observations at multiple frequencies measure the
thermal emission associated with H\,{\sc ii} regions and non-thermal
emission related to supernova remnants \citep[e.g.,][]{turner83,
  johnson04}.  Radio recombination lines are fainter than radio
continuum emission, but they directly measure the thermal emission and
yield important information about the dynamics of these star clusters.
\citet{anantharamaiah00} combined both RRL and continuum data at
several different frequencies to constrain models of these young
massive star clusters.  Technical improvements provided by the JVLA
increased the sensitivity of such work by almost an order of magnitude
\citep{kepley11}, and enabled RRL studies of less massive star
clusters.  For example, \citet{balser17} detected RRL emission in
IC\,342 for the first time and revealed thermal emission to the east
and west of the nuclear star cluster that was associated with giant
molecular clouds (see Figure~\ref{fig:ic342}).  The best fit model is
a collection of many hundreds of compact ($\sim 0.1\,$pc) H\,{\sc ii}
regions ionized by an equivalent of $\sim 2000$ O6 stars.  Even with
the increased sensitivity of the JVLA these observations are
challenging and we do not detect RRL emission in gas located beyond
the most central regions.  Better sensitivity over many spatial scales
is required.  Also, \citet{anantharamaiah00} showed that because of
free-free opacity effects the RRL are brighter at higher frequencies
and therefore it is important to observe RRLs over a large range of
frequencies (e.g., 1-100\,GHz).

The ngVLA has the required sensitivity spanning a wide range of
spatial and frequency scales.  Ideally, we want to resolve individual
H\,{\sc ii} regions but also have the sensitivity to detect extended
emission (e.g., the diffuse ionized gas).  So we need to probe scales
from $\sim 1-1000\,$pc in galaxies with distances between 1-20\,Mpc,
or angular scales 0$\rlap.{^{\prime\prime}}01 - 200^{\prime\prime}$.
To estimate the required spectral sensitivity we use a model H\,{\sc
  ii} region.  We assume a homogeneous sphere ionized by an O5 star at
a distance of 2\,Mpc with an electron density and temperature of
1,000\,cm$^{-3}$ and 8,000\,K, respectively.  Using the stellar models
of \citet{martins05} the continuum flux density at a frequency of
5\,GHz is 0.1\,mJy \citep[see appendix G.1 in][]{balser95}.  This
continuum flux density estimate is similar to the values found in
compact sources in nearby galaxies \citep[e.g.,][]{tsai06}.  Assuming
a line-to-continuum ratio of 0.1 yields a RRL intensity of
$10\,\mu$Jy.  Using the ngVLA sensitivity from \citet{selina17} for a
50\,mas beam, we estimate an integration time of 35\,hr assuming a SNR
of 5, a spectral resolution of 5\,km\,s$^{-1}$, and averaging 20 RRLs.
This is over two orders of magnitude more sensitive than our JVLA RRL
observations.  For an optically thin, unresolved H\,{\sc ii} region
the results are not very sensitive to the electron density, electron
temperature, or frequency.  Moreover, the flux density only increases
by approximately a factor of two for an O3 star relative to an O6
star.  The integration time can therefore be approximated as $t_{\rm
  intg} \approx 35*(D/2\,{\rm Mpc})^{4}\,$hr for detecting RRL
emission from H\,{\sc ii} regions ionized by a single star, where $D$
is the distance.

The RRL data probe the ionized gas and will complement observations
that sample the neutral (H\,{\sc i}; ngVLA) and molecular (CO; ALMA,
ngVLA) components to characterize these star formation complexes.
Detection of submillimeter H and He RRLs with ALMA trace the hardness
of the radiation field, or indirectly the high-mass end of the initial
mass function, in nearby massive young star clusters \citep[e.g.,
see][]{scoville13}.  Similar observations with the ngVLA are possible
at 3\,mm assuming the ngVLA is about five times more sensitive than
ALMA (A. Bolatto, private comm.).  In principle the JWST Pa$\alpha$
and ngVLA RRL intensity ratio could directly measure the dust
extinction as a function of velocity.

\vskip5pt

\section{Summary}

The ngVLA will allow for sensitive observations of RRLs and impact the
following areas:

\begin{itemize}

\item {\it Galactic Structure:} The ngVLA will have the sensitivity to
  create a Galaxy-wide, volume-limited sample of H\,{\sc ii} regions
  for the first time.  This is needed to determine the global star
  formation properties of the Milky Way, and to better characterize
  both the morphological and chemical structure of the Galactic disk.

\item {\it H\,}\textit{\fauxsc{ii}} {\it Region Physics:} The ngVLA
  could solve several long standing problems in H\,{\sc ii} region
  physics.  These include discriminating between different models of
  the collision rates from free electrons that determine the opacity,
  characterizing temperature fluctuations which effect abundance
  determinations, and directly measuring the magnetic field strength
  within H\,{\sc ii} regions.

\item {\it Star Formation in Nearby Galaxies:} The ngVLA will have
  both the necessary sensitivity and spatial resolution to probe star
  formation in a statistically significant sample of nearby galaxies.
  RRLs provide an extinction-free tracer of the physical conditions
  and dynamics of the ionized gas.

\end{itemize}




\end{document}